\documentclass[fleqn,10pt]{wlscirep}
\usepackage{graphicx,amsmath,amssymb,wasysym,verbatim,color}
\newcommand{\be}{\begin{equation}}
\newcommand{\ee}{\end{equation}}

\title{Fragmentation transitions in a coevolving nonlinear voter model}
\author[1,$*$]{Byungjoon Min}
\author[1,$\dagger$]{Maxi San Miguel}
\affil[1]{IFISC, Instituto de F\'isica Interdisciplinar y Sistemas Complejos
(CSIC-UIB), Campus Universitat Illes Balears, E-07122 Palma, Spain}
\affil[*]{byungjoon@ifisc.uib-csic.es}
\affil[$\dagger$]{maxi@ifisc.uib-csic.es}

\begin{abstract}
We study a coevolving nonlinear voter model describing the coupled 
evolution of the states of the nodes and the network topology. 
Nonlinearity of the interaction is measured by a parameter $q$. 
The network topology changes by rewiring links at a rate $p$. 
By analytical and numerical analysis we obtain a phase diagram in 
$p,q$ parameter space with three different phases: Dynamically active 
coexistence phase in a single component network, absorbing consensus 
phase in a single component network, and absorbing phase in a 
fragmented network. For finite systems the active phase has a lifetime 
that grows exponentially with system size, at variance with the similar 
phase for the linear voter model that has a lifetime proportional to system 
size. We find three transition lines that meet at the point of the 
fragmentation transition of the linear voter model. A first transition line 
corresponds to a continuous absorbing transition between the active and 
fragmented phases. The other two transition lines are discontinuous 
transitions fundamentally different from the transition of the linear 
voter model. One is a fragmentation transition between the consensus and 
fragmented phases, and the other is an absorbing transition in a single 
component network between the active and consensus phases.
\end{abstract}

\begin{document}
\flushbottom
\maketitle
\thispagestyle{empty}

\section*{Introduction}
The structure of networks on which interactions between individuals
take place affects the dynamical processes between agents \cite{boccaletti,barrat}.
At the same time, the pattern of social ties constantly changes under the strong
influence of the state of individuals \cite{mcpherson,centola,ba,bmin}.
Coevolution of individual states and network structures \cite{coevol0}
in a comparable time scale is commonly observed
in reality \cite{gross-review,gross-book}.
For example, individuals who are connected on a given network may
become alike since they interact with each other via an existing social tie.
At the same time social connections can also be established
between people because they have some similarity \cite{mcpherson}.
In order to model the coevolutionary dynamics, there have
been several attempts combining dynamics on top of networks and
evolution of networks according to the state of individuals 
in different contexts of game theory~\cite{zimmerman,pacheco,szolnoki,perc,wang},
opinion formation~\cite{vazquez,vazquez2,herrera,holme}, and
epidemic dynamics~\cite{gross-prl,coevol,scarpino} among other.
One of the simplest yet fruitful models is a coevolving voter
model, describing the change of the state of the nodes following the rule of the
voter model dynamics and organization of a network by rewiring
links~\cite{vazquez,holme,durrett,kimura,bohme,demirel,diakonova1,diakonova2,klimek}.
The coevolving voter model exhibits a fragmentation transition from a single
connected network to a network with two components \cite{vazquez,holme}.

In the original voter model, a voter (a node of the network) can be in either of two
discrete states and it imitates the state of one of its neighbors chosen randomly.
This simple rule implies a dyadic interaction in which the state of the majority of the
neighbors does not play a role other than in an average manner. However, several
neighbors can influence collectively an individual's state so that an agent may
consider the state of more than one neighbor to change its state, rather than blindly
copying the state of one of its random neighbors. This implies a nonlinear
interaction of an agent with its neighborhood implemented as a nonlinear voter
model \cite{castellano,schweitzer}. Conceptually, the difference between the ordinary
voter model and the nonlinear voter model is the same discussed for simple and complex
contagion processes \cite{contagion,czaplicak}.
The same form of nonlinear interaction with the
neighborhood of the nonlinear voter model is the one considered to model social
pressure in social impact theory \cite{nowak} as well as in language competition
dynamics, named volatility \cite{abrams,jstat}, or in language evolution problems\cite{nettle}.

In this paper, we study the role of nonlinearity in a coevolving
voter model by combining evolutionary dynamics of networks and
a nonlinear voter model characterized by a degree of nonlinearity $q$.
To be specific, we update the state or the links of node $i$ in the network with
a probability $\rho_i^q$ where $\rho_i$ is the fraction of neighbors of $i$ 
in a different state than $i$.
Mathematically $\rho_i$ is defined as $\frac{a_i}{k_i}$ where $k_i$ is the
total number of neighbors called the degree of $i$ and $a_i$ is the number 
of active links connecting $i$ to a node in a different state.
When $q$ is larger than $1$, an agent
follows the majority opinion more frequently than in the ordinary voter model.
By contrast, minority opinion of neighbors is more likely to be
chosen than in the linear model when $q$ is less than $1$.
When the parameter $q$ takes integer values, it can be interpreted as a voter selecting
multiple neighbors and changing its state when all selected neighbors have
unanimously a different state \cite{castellano}. However, $q$ is taken as a continuous
parameter in social impact theory or in language evolution dynamics. While it is argued
that $q<1$ in social impact theory \cite{nowak} and  language evolution \cite{nettle},
it was fitted to $q=1.3$ to account for data on different language extinction
processes \cite{abrams}. We study the role of nonlinearity measured by $q$
in the fragmentation transition of the voter model:
We find that the coevolving nonlinear voter model still shows a fragmentation transition
between connected and disconnected networks but with different mechanisms
depending on the nonlinearity $q$. For $q<1$ the system undergoes a continuous
transition between a dynamically active phase and an absorbing frozen state with two
disconnected clusters each of them in a different consensus state, i.e. in each cluster
all the nodes are in the same state, but this state is different in both clusters.
This is the same transition found for the ordinary voter model ($q=1$)
in the thermodynamic limit $N \to \infty$.
However, for finite but large systems $N \gg 1$ the coevolving linear voter model, $q=1$,
reaches a frozen state either in a fully consensus or a fragmented phase
while the nonlinear coevolving model with $q<1$ can remain in a dynamically active
phase for observation times that grow exponentially with $N$.
However, for $q>1$ the transition is abrupt and between two absorbing states,
therefore essentially different than the continuous transition observed in the ordinary
voter model.

\section*{Results}
\subsection*{Coevolving nonlinear voter model}
\begin{figure}[t]
\includegraphics[width=0.55\linewidth]{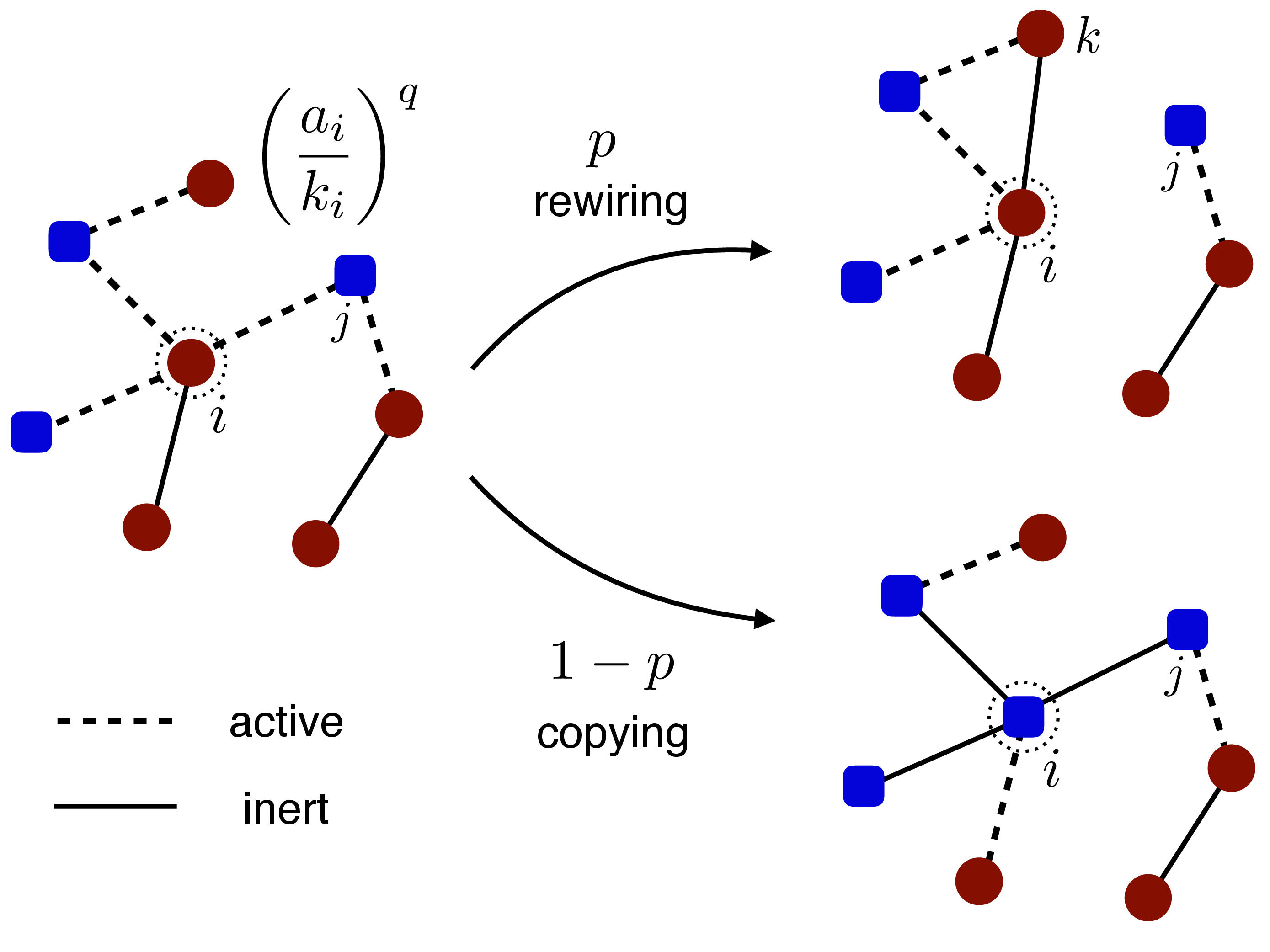}
\caption{
Schematic illustration of update rule of the coevolving nonlinear voter model.
Each node is in either up (red circle) or down (blue rounded square) state.
Solid and dashed lines indicate respectively inert and active links.
At each step, we randomly choose a node $i$.
And we choose one of its neighbors $j$ connected by an active link
with a probability $\left( \frac{a_i}{k_i} \right)^q$.
Then, we rewire an active link with a probability $p$
and copy the state of the neighbor with a probability $1-p$.
}
\label{fig:model}
\end{figure}

We consider a degree regular network in which each node has
the same number $\langle k \rangle$ of random neighbors.
Each node $i$ is initially either in a state $s_i=+1$ (up) or $-1$ (down)
with the same probability $1/2$.
In a given configuration, each link can be classified into two
different types, active or inert. We define active (inert) links for
the links connecting two nodes in different (same) states.
We also define the density of active links $\rho_i$ for each node $i$
as $\frac{a_i}{k_i}$ where $k_i$ is the degree and $a_i$ is the number
of active links of node $i$.
At each step, we randomly choose a node $i$. And with a probability $\rho_i^q$,
we choose at random an active link to a neighbor, say $j$.
Note that with the complementary probability $1-\rho_i^q$, nothing happens
and we pick another node randomly.
Once we choose node $i$ and $j$, with a probability $p$ node $i$ removes
the link to $j$ and rewires a link to another node, say $l$, having
the same state with $i$.
And with a probability $1-p$, node $i$ flips its state to become the
same as the state of node $j$ (see Fig.~\ref{fig:model}).
This proceeds, keeping the average degree $\langle k \rangle$ constant, until the
system reaches a dynamically active steady state or an absorbing configuration.

The probability $p$, or plasticity parameter, represents a ratio of the time scale
of evolution of the network and the time scale of evolution of the states of the
nodes. When $p=0$, the model corresponds to a
nonlinear voter model on a static network. In contrast, for $p=1$, the model
describes a rewiring process to become two differently ordered groups.
The degree of nonlinearity $q$ controls the frequency of the selection of
agents for execution. When $q>1$, nodes with more active links have
a higher chance to be changed than the ordinary voter model.
In contrast, if $q<1$, nodes with less active links are more likely to be changed.
The model becomes the original voter model when $q=1$.

\subsection*{Coexistence, consensus, and fragmented phases}

\begin{figure}[t]
\includegraphics[width=0.6\linewidth]{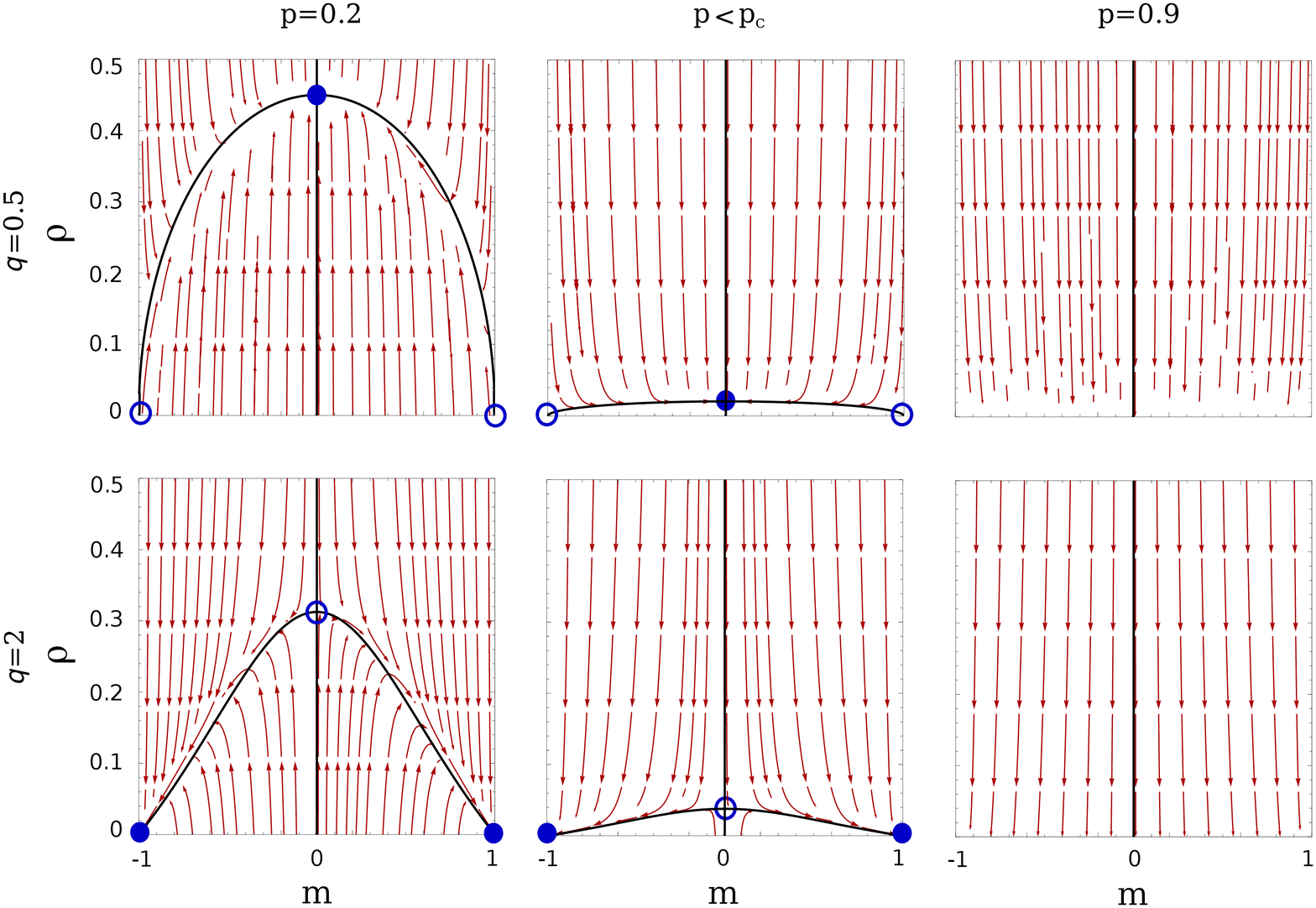}
\caption{
Flow diagram on the plane $(m, \rho)$ for $q=0.5$ and $2$
below and above $p_c = 0.875$ (q=0.5) and $0.8$ (q=2)
obtained from pair approximation.
The filled (open) circles denote
the stable (unstable) fixed points. The point at $m=0$
is stable for $q=0.5$ but is unstable for $q=2$.
}
\label{fig:theory}
\end{figure}

The magnetization defined as $m=(1/N)\sum_i s_i$ and the density of active links
$\rho$ over a network satisfy coupled equations of evolution derived in a
pair approximation in the Methods Section. The steady state solutions of
these equations are:
$(m,\rho)=(-1,0)$, $(1,0)$, $(0,\rho^*)$ and $(m^*,0)$.
The solutions $(-1,0)$ and $(1,0)$ represent a consensus absorbing frozen
phase with all nodes in the same state. In contrast, $(0,\rho^*)$ indicates
a dynamically active phase with coexistence of the same number of nodes in
the  up and down states where $\rho^*$ represents the density of
active links of the active phase. For the steady state solution with $m=0$,
the equation of $\rho$ (Eq.~\ref{eq:mf2}) reduces to
\begin{eqnarray}
\rho^{q}\left\{ -p + (1-p)[\langle k \rangle - 2 q -2 (k-q)\rho] \right\}=0,
\end{eqnarray}
and $\rho^*$ is obtained as
\begin{eqnarray}
\label{eq:rho}
\rho^*=\frac{(1-p)(\langle k \rangle - 2 q) -p}{2(1-p)(\langle k \rangle -q)}.
\end{eqnarray}
Note that when $q=1$, we recover the result for the coevolving linear voter
model~\cite{vazquez}, $\rho=\frac{(1-p)(\langle k \rangle -1) -1}{2(1-p)(\langle k \rangle -1)}$.
Finally, the $(m^*,0)$ solution corresponds to a fragmentation of the network
in two components, each of them separately ordered.
In general $m^*$ is determined by the initial fraction of up and down nodes
and in particular $m^*=0$ for our initial condition $m=0$.
For different initial conditions, i.e. $m\ne 0$, the fragmented components
can have different sizes.
Note that solutions with $\rho=0$ correspond to absorbing states in which the
dynamics is frozen, including the fragmented phase  $(m^*,0)$.

In summary, there are three different types of stationary solutions with
the initial condition $m=0$:
(i) $(m,\rho)$=$(0,\rho^*)$ corresponding to a dynamically active phase of coexistence in
a single component network,
(ii) $(-1,0)$ or $(1,0)$ which are consensus absorbing phases in a single
component network, and
(iii) $(0,0)$ corresponding to an absorbing fragmented phase with a network
composed of two disconnected clusters.

When $p=0$, that is a nonlinear voter model on static networks,
Eqs.~\ref{eq:mf1} and \ref{eq:mf2} have three sets of solutions,
$(-1,0)$, $(1,0)$, and $(0,\rho^*)$, while the solution $(0,0)$ corresponding to
a fragmentation solution exists for $p\neq0$.
For small values of $p$, $(0,0)$ is unstable, becoming stable by increasing $p$
at a point $p_c$ where $\rho^*=0$ in Eq.~\ref{eq:rho}. Explicitly, the transition
point is given by $p_c=\frac{\langle k \rangle -2 q}{1+\langle k \rangle-2q}$
in the pair approximation. When $p>p_c$, the fragmented phase $(0,0)$ is
stable solution for all value of $q$.
When $p<p_c$, the system is found in the dynamically active or absorbing phase
in a single component network depending on the stability of the solution $(0,\rho^*)$ which
is determined by the slope of Eq.~\ref{eq:mf1} at a given $q$. If $q<1$, the
solution $(0,\rho^*)$ is stable, so the dynamically active phase is predicted.
But when $q>1$, the solution becomes unstable and $(-1,0)$ and $(1,0)$
are stable, so that the absorbing consensus phase is predicted.

In Fig.~\ref{fig:theory}, the dynamical flow on the plane $(m,\rho)$ to stable
fixed points and the stability of them predicted from
Eqs.~\ref{eq:mf1} and \ref{eq:mf2} are depicted.
The system evolves by following the flow from initial state and
eventually reaches stable fixed points (filled circle) or $\rho=0$
line corresponding to an absorbing phase.
For the case $q=0.5$, when $p<p_c$, $\rho^*$ at $m=0$ is stable
indicating coexistence of the two possible states of the nodes in a single connected network.
The value of the number of active links $\rho^*$ in the stable point gradually
decreases with increasing $p$ and becomes zero at $p_c$.
When $p>p_c$ and for the initial condition $m=0$, the system eventually
arrives at the absorbing state $\rho=0$ and $m=0$ implying a network with two
disconnected clusters with a different consensus in each of the clusters.
Therefore, the mean-field analysis predicts a continuous transition
between the active phase $(m,\rho)$=$(0,\rho^*)$ and an absorbing phase
with two split clusters $(0,0)$.

However, when $q=2$, the stable fixed points are located at $(-1,0)$ and $(1,0)$
and the solution at $m=0$ is unstable for $p<p_c$.
Above $p_c$, $\rho=0$ and $m=0$ (for initial condition $m=0$) is the steady state
corresponding to the absorbing phase with a fragmented network.
Therefore, at the transition the magnetization drops abruptly to be zero
at $p_c$ in a discontinuous transition.
Thus we predict different phase transitions
depending on $q$, that is a continuous phase transition from an active to a frozen absorbing state
for $q\le 1$ and a discontinuous transition between two absorbing states for $q>1$.

\begin{figure}[t]
\includegraphics[width=0.6\linewidth]{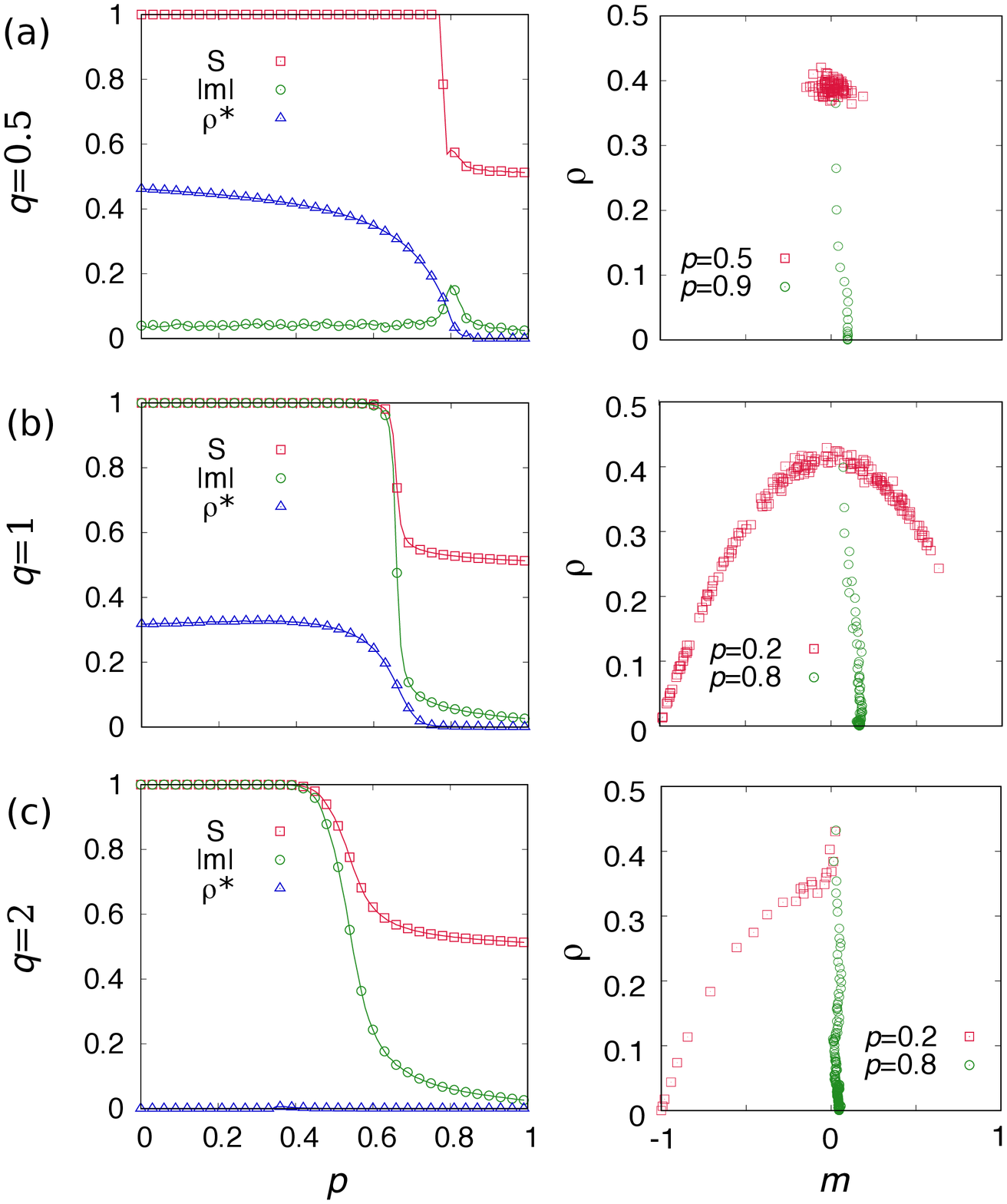}
\caption{
The size $S$ of giant component, magnetization $|m|$,
and the density $\rho$ of active link for the network at the steady state
for (a) $q=0.5$, (b) $q=1$, and (c) $q=2$ on random regular networks
with $\langle k \rangle =8$, $N=10^3$, and initial condition $m=0$
averaged over $10^4$ realizations.	
The right panel represents an typical trajectory to steady state below
and above $p_c$ on $(m,\rho)$ space.
}
\label{fig:three}
\end{figure}

\subsection*{Numerical results and finite-size analysis}
In order to examine the different mechanisms of the fragmentation
transition, we numerically measure the size of the largest component $S$,
absolute value of magnetization $|m|$,
and the density of active links for the network
$\rho^*$ at the steady state for $q=0.5$, $1$, and $2$ (Fig.~\ref{fig:three}).
The fragmented states above $p_c$ for the three values of $q$ share the common
feature characterized by $(S, m, \rho^*)$ $\approx$ $(0.5, 0, 0)$,
indicating a network with two disconnected clusters
each of them in a different consensus state.

Below $p_c$, the network is in a single connected component regardless
of the value of $q$ but differently characterized
by  $(1, 0, \rho^*_{0.5})$ for $q=0.5$, $(1, 1, \rho^*_{1})$ for $q=1$,
and $(1, 1, 0)$ for $q=2$, where $\rho^*$ is a non-zero value.
For $q=0.5$, the network forms a single component with a certain fraction
of active links meaning that up and down states coexist.
So, the system, in the thermodynamic limit, remains in a dynamically active
state and does not reach the absorbing state. The value shown for $\rho^*$
corresponds to the one in this thermodynamic limit and it is measured
numerically as the density of active links averaged over surviving runs.
On the contrary, when $q>1$, the system reaches the consensus phase
showing all nodes are in the same state within a single component.
For $q=1$ which is the linear coevolving voter model, the active state
is observed by showing non-zero $\rho^*$ as expected in the analytic
prediction, Eqs.~\ref{eq:mf1} and \ref{eq:mf2}. $\rho^*$ is again calculated
averaging only over surviving runs. However, due to finite size fluctuations
the system goes to an absorbing consensus phase
below $p_c$ \cite{vazquez} showing $m=1$ in Fig.~\ref{fig:three}(b).

In the right panels in Fig.~\ref{fig:three}, a typical trajectory on $(m, \rho)$ space
also exhibits different mechanisms to approach the steady state depending on the value of $q$.
Red symbols ($\square$) for $q=0.5$ remain around $\rho \approx 0.4$ indicating the active
state but those for $q=2$ collapses into the absorbing state directly $(m=-1, \rho=0)$.
When $p>p_c$, the system reaches the fragmented phase for all values of $q$,
as shown by the green symbols ($\diamond$). Deviation from zero in the final state of
magnetization $m$ in a given trajectory is originated by finite size fluctuations.

\begin{figure}[t]
\includegraphics[width=0.6\linewidth]{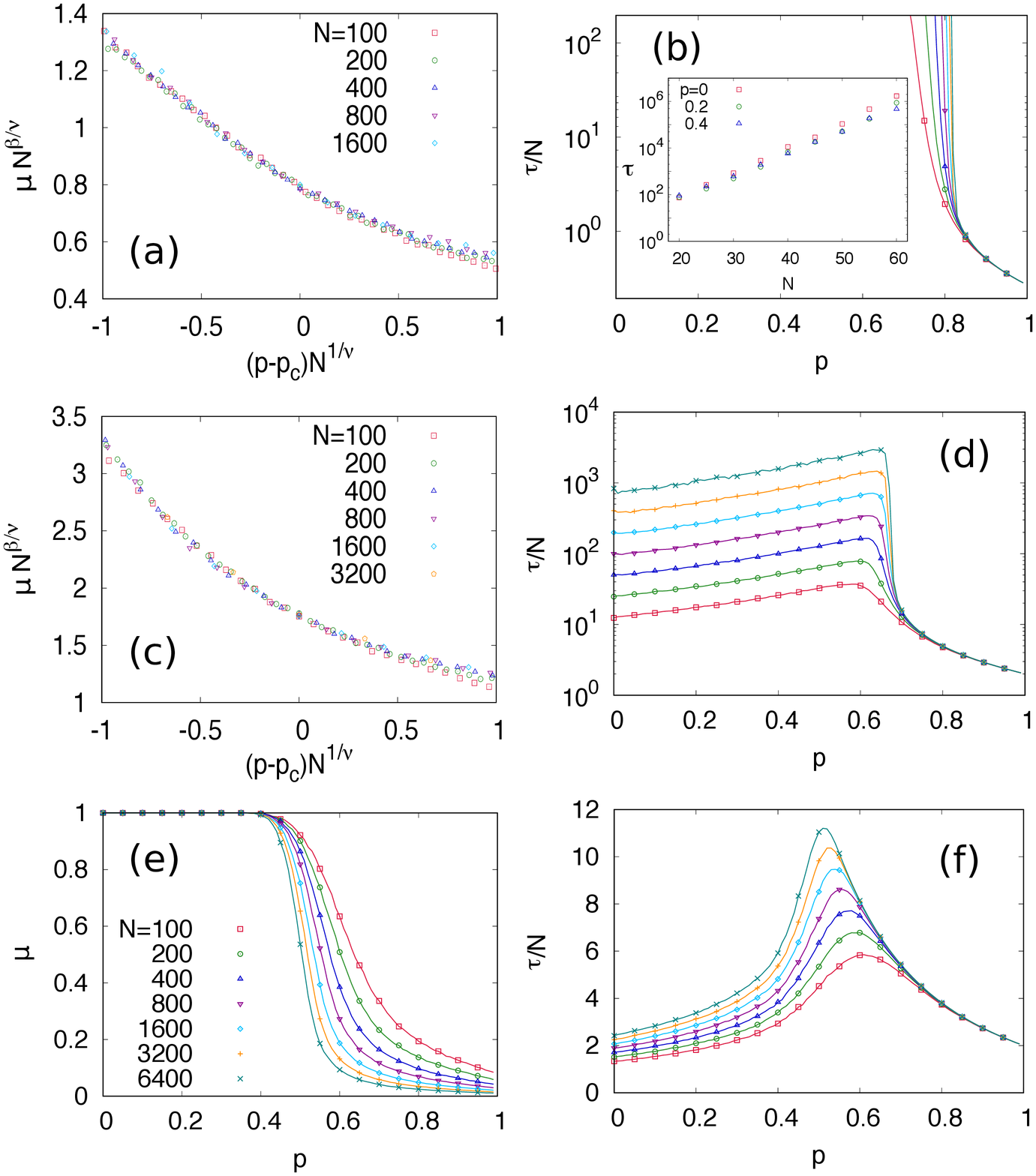}
\caption{
Rescaled magnetization $\mu$ and time to steady state state $\tau$ for
(a,b) $q=0.5$, (c,d) $q=1$, and (e,f) $q=2$ with different network size
$N$ and $\langle k\rangle=8$, averaged over $10^5$ runs.
Figures (a) $q=0.5$ and (c) $q=1$ show scaling of magnetization in
a form $\mu=N^{\beta/\nu} f\left(N^{1/\nu}(p-p_c)\right)$
with $p_c\approx 0.83$ and $0.68$, respectively.
When $q=0.5$, (b) $\tau$ exponentially grows with respect to $N$ below $p_c$.
For $q=2$, (e) the transition of magnetization is getting sharp as $N$ increases,
with (f) non-diverging $\tau$ at the transition.
}
\label{fig:finite}
\end{figure}

The differences between a continuous or discontinuous phase transition
depending on the nonlinearity are illustrated in Fig.~\ref{fig:finite}
for different system sizes $N$. We assume that a scaling relation has a form
\begin{equation}
\mu=N^{\beta/\nu} f\left(N^{1/\nu}(p-p_c)\right)
\end{equation}
where $\mu$ is a rescaled magnetization, defined as $\mu=|m|$,
so that $\mu$ is $1$ when the system is fully ordered either up or down
and $\mu$ is zero when disordered.
We find that the magnetization shows a continuous phase transition
with critical exponents $\beta/\nu=0.319(1)$ for $q=0.5$ 
and $\beta/\nu=0.317(2)$ for $q=1$ and a diverging time to consensus at
criticality. 
It is important here to note a singularity on the time to reach the
absorbing state: while it scales linearly with $N$, $\tau \sim N$, for
$q=1$ \cite{vazquez}, this time increases exponentially $\tau \sim e^N$
for $q<1$. In practice this means that for $p<p_c$ and $q<1$ the system
remains in the active dynamically state for normal observation times of
a large system, at variance with the case $p=1$. For example in the
trajectories shown in the right panel of Fig. 3 for $p<p_c$, in the same
time scale the system has reached the absorbing state of $q=1$ but it
remains around the nonzero initial value of $\rho$ for $q=0.5$.
When $q=2$, the transition becomes sharp at $p_c \approx 0.47$ as $N$ increases
and it is expected to be discontinuous in the limit $N \to \infty$.
The discontinuous transition for $q=2$ is supported by
non-diverging consensus time $\tau$ and the pair approximation analysis as well.

\subsection*{Phase diagram}
\begin{figure}[t]
\includegraphics[width=0.8\linewidth]{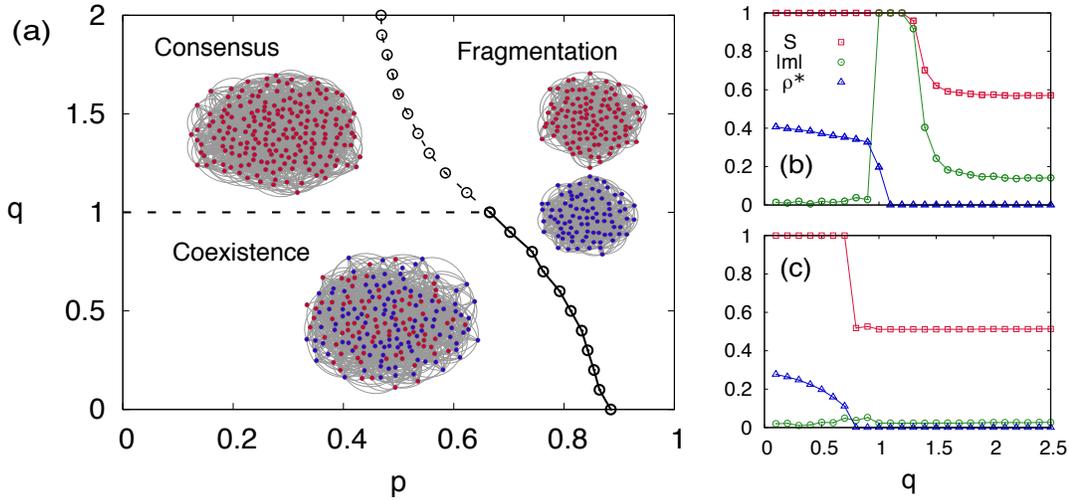}
\caption{
(a) Phase diagram with respect to $p$ and $q$ shows consensus, coexistence,
and fragmented phases, obtained numerically on degree regular networks with
$\langle k\rangle=8$, $N=10^4$ and initial condition $m=0$,
averaged over $10^3$ realizations.	
Examples of	network configuration at the steady-state of the coevolution model
are also shown with $N=200$ and $(p,q)=(0.2,0.5)$ for coexistence,
$(0.2,2)$ for consensus, and $(0.8,0.5)$ for fragmentation.
Size of giant component, magnetization, and density of active links
at (b) $p=0.55$ and (c) $p=0.75$ are also shown.
}
\label{fig:network}
\end{figure}

We finally examine numerically the phase diagram of the system with respect to both
parameters, plasticity $p$ and degree of nonlinearity $q$ and
also the fragmentation transition when varying $q$. Fig~\ref{fig:network} (a) shows
the phase diagram indicating consensus, coexistence, and fragmentation phases
for different $p$ and $q$.
Typical examples of a network in each phase above and below $p_c$ are depicted in
Fig.~\ref{fig:network} (a). When $p>p_c$,
the rewiring process is dominant, so the network is divided into two clusters.
The two clusters are in opposite states and each cluster is in a full consensus.
Therefore, when $p>p_c$, the system both for $q<1$ and $q>1$ reaches an absorbing state,
with two oppositely ordered clusters.
However, the steady state of the coevolving model below $p_c$
relies on the degree of the nonlinearity $q$.
If $q>1$, all nodes belong to a single component with the same state.
On the other hand, for the case $q<1$, the system does not reach an
absorbing state and remains in an active state containing both up and
down nodes in the network.

By varying the degree of nonlinearity $q$, a discontinuous transition between
consensus and coexistence phases occurs at $q=1$ for low $p$ regime.
On the other hand, as $p$ increases, a fragmented phase appears from
either the consensus ($q>1$) or coexistence ($q<1$) phases. Therefore, for
high enough $p$ only the fragmented phase exists for any value of $q$.
At intermediate values of $p$, three phases can be observed depending
on the nonlinearity $q$. For example, at $p=0.55$ we find two transitions
from coexistence to consensus and subsequently from consensus to the fragmented
phase as shown in Fig.~\ref{fig:network} (b). We also observe a direct
transition between coexistence and fragment phases as shown in
Fig.~\ref{fig:network} (c) at $p=0.75$.
In summary, for $p<p_c(q)$ a discontinuous transition exists at $q=1$ from a
dynamically active state to a frozen absorbing state of consensus in a single
component network. At $p_c(q)$ there is a fragmentation transition to an absorbing
state in which the network is fragmented in two components, each of them in
a different consensus state,

\section*{Discussion}
In this paper, we have examined the role of nonlinearity in the coevolving voter
model. We find diverse phases depending on the nonlinearity $q$ and the rate
of rewiring or plasticity parameter $p$. We also find that the mechanisms of
network fragmentation are different for the cases $q<1$ and $q>1$. When $q<1$,
the system undergoes an absorbing phase transition at $p=p_c$ between one connected component
in an active state and two disconnected clusters. In addition, the transition is
continuous and characterized by the density of active links $\rho$ vanishing at
the transition point. This is the same type of transition observed for the linear
voter model ($q=1$), but the time to reach the absorbing state by finite size
fluctuations in the connected phase which scales linearly with $N$ for $q=1$,
scales now exponentially with $N$ for $q<1$. 
When $q>1$, the system reaches an absorbing frozen
state with $\rho=0$ for any $p$. For $p<p_c$, the network is in a full consensus
within a single component, while for $p>p_c$ the network fragments with a
discontinuous transition at $p=p_c$. This is a different type of transition to
the one found in the linear voter model. Therefore, our results reveal that the nonlinearity
modifies nontrivially the fragmentation
transition known for the ordinary voter model $(q=1,p=p_c(q=1))$,
which is a special case at the boundary between two different fragmentation transitions.
In fact, three transitions lines meet in this
point in the $(q,\/ p)$ parameter space: A continuous
fragmentation transition $(q<1, p=p_c)$, a discontinuous fragmentation
transition $(q>1, p=p_c)$ and an absorbing transition from an active to a frozen
phase in a single component network $(q=1, p<p_c)$.

\section*{Methods}
\subsection*{Mean-field approximation of nonlinear voter model}
We first examine the nonlinear voter model on fully-connected networks where
the rewiring process is not well defined,
so that we take $p=0$.
A mean-field equation of the magnetization defined as
$m=(1/N)\sum_i s_i$ for the fully connected networks in the limit
$N\to \infty$ is simply given by
\begin{align}
\frac{d m}{d t}= 2 \left[ -\left( \frac{1+m}{2} \right) \left( \frac{1-m}{2} \right)^q +\left( \frac{1-m}{2} \right) \left( \frac{1+m}{2}\right)^q \right].
\end{align}
Once $m$ is given, the density of active links is determined by
\begin{align}
\rho =\frac{1}{2}(1-m^2).
\end{align}
There are three steady state solutions of eq. (4), $m=-1$, $0$, and $1$ for any $q$
The stability of the solutions depends on $q$.
If $q<1$, the neutral state $m=0$ showing the same fraction of up
and down states is stable.
However if $q>1$, the absorbing states either $m=-1$ or $m=1$ become
stable, so the system is in a fully ordered state.
For $q=1$, the magnetization $m$ is conserved as known for the ordinary voter model.

\subsection*{Pair approximation of coevolving nonlinear voter model}
In a random network, the density of active links $\rho$ and the
magnetization $m$ are coupled quantities. We study their coupled evolution
within a pair approximation for well mixed populations.
We assume degree regular random networks. Due to the evolution of networks
($p\neq0$), the degree regular structure is not maintained in time, but the
assumption is still reasonable because an homogeneous structure
is observed due to the random rewiring process.
Given that we pick a node with state $s$, the conditional probability that
we select a node connected to it, but in a
different state is given by $\rho/(2 n_{s})$ where
$s\in \{+,-\}$ and $n_{s}$ represents the global fraction of nodes in $s$ state.
In addition, when we choose a node with state $s$ to be updated with
the probability $[\rho/(2 n_s)]^q$, this node has approximately $q$ neighbors
in a different state. And, the other neighbors $k-q$
have  a different state with probability $\rho/(2 n_{s})$.
Putting all these together, the coupled equations for $\rho$
and $m$ in a mean-field pair approximation level with
$N\to \infty$ \cite{vazquez,demirel,vazquez3,mf} are:
\begin{align}
\label{eq:mf1}
\frac{dm}{dt}=&2(1-p)\left[ - n_{+} \left( \frac{\rho}{2 n_{+}} \right)^q + n_{-}\left( \frac{\rho}{2 n_{-}} \right)^q \right],\\
\label{eq:mf2}
\frac{d \rho}{dt}= &\frac{2}{\langle k \rangle} \Bigg\{ -p\left[ n_+ \left( \frac{\rho}{2n_{+}} \right)^q +n_{-}	 \left( \frac{\rho}{2 n_{-} }\right)^q \right] +(1-p)\left[ n_{+} \left(\frac{\rho}{2 n_{+}}\right)^q \left(\langle k\rangle -2 q -2 (\langle k \rangle -q) \frac{\rho}{2 n_{+}} \right)\right] \\
&+(1-p)\left[ n_{-} \left(\frac{\rho}{2 n_{-}}\right)^q \left(\langle k\rangle -2 q -2 (\langle k \rangle -q) \frac{\rho}{2 n_{-}} \right)\right] \Bigg\} \nonumber,
\end{align}
where $n_+=(1+m)/2$ and $n_-=(1-m)/2$.
The steady state solutions of these equations and their stability are discussed in the main text.

\section*{Acknowledgements}
We thank M. Cosenza for useful discussions. This work was supported by the
Spanish  Ministry  MINEiCO and FEDER (EU) under the project ESOTECOS (FIS2015-63628-C2-2-R).

\section*{Author Contributions}
All authors conceived the study, performed the research, analyzed data, and wrote the paper.

\section*{Additional Information}
Competing Interests: The authors declare no competing financial interests.

\end{document}